\documentclass[a4paper]{jpconf}

\usepackage{graphicx}
\usepackage{amsmath}
\usepackage{amssymb}
\usepackage{amsfonts}
\usepackage[usenames,dvipsnames]{color}

\newcommand{\fourmat}[4]{{\begin{pmatrix}
                         {#1} & {#2} \\ {#3} & {#4}
                         \end{pmatrix}}}
\newcommand{\twospinor}[2]{{\begin{pmatrix}
                           {#1} \\ {#2}
                           \end{pmatrix}}}

\newcommand{\nuc}[2]{$^{#1}${#2}}

\renewcommand{\vec}[1]{{\textbf{#1}}}

\begin{document}

\title{Structure of superheavy nuclei}

\author{Michael Bender}

\address{Univ. Bordeaux, CENBG, UMR5797, F-33175 Gradignan, France, and}
\address{CNRS/IN2P3, CENBG, UMR5797, F-33175 Gradignan, France}

\ead{bender@cenbg.in2p3.fr}

\author{Paul-Henri~Heenen}
\address{Physique Nucl\'eaire Th\'eorique,
         Universit\'e Libre de Bruxelles, C.P. 229, B-1050 Bruxelles,
         Belgium}

%
%

\begin{abstract}
Properties of transactinide nuclei needed 
for the interpretation of spectroscopic data and for the 
extrapolation to the expected "island of stability" of 
superheavy nuclei are reviewed from a theorist's point
of view in the framework of self-consistent mean-field models.

\end{abstract}

%
%

\section{Introduction}

Superheavy elements (SHE) are usually defined as the chemical elements
whose atomic nuclei have a vanishing fission barrier in the liquid drop
model. In this schematic picture, for nuclei with $Z \gtrsim 106$
the strong Coulomb repulsion pushes the nucleus to large deformation
and overcomes the surface tension that tries to keep the nucleus at 
small deformation. That such systems can exist and have been indeed observed 
is due to quantal shell effects: according to Strutinski's theorem, a less 
than average density of single-particle levels around the Fermi energy 
gives additional binding, whereas a larger than average value reduces 
the binding. As a consequence, the variation of the single-particle 
spectrum with deformation creates local pockets and barriers in the
deformation energy surface. Being on a scale of about 10~MeV, the amplitude 
of these variations is small compared to the total binding energy
of almost 2 GeV, but it is large enough to stabilise many heavy nuclei
against spontaneous fission. $\alpha$-decay, $\beta$-decay or electron 
capture often become their dominating decay modes instead. SHE are also 
often called \emph{transactinides}. 

Over the last four decades, the limits of the chart of nuclei 
have been pushed far into the realm of superheavy elements up to 
$Z=118$ \cite{Hof98a,Arm99a,Hof00a,Arm00a,Oga07a,Hof11a,Hee09a}. 
One motivation for the experimental efforts in this field of nuclear research 
is the long-standing prediction of an island of very long-living spherical 
SHE, which originally was expected to be centred at $Z \approx 114$ and 
$N \approx 184$ \cite{Nil68a,Mel69a,Mos69a,Koh71a,Sob71a}.
Even when the expectations about their lifetimes
had to be cut down from years or even longer to minutes, 
perhaps just seconds, finding such nuclei would still provide a unique 
laboratory to study the nuclear many-body system under extreme conditions.

Because of the quickly dropping fusion cross sections 
\cite{Hof98a,Arm99a,Hof00a,Arm00a,Oga07a,Zag12a}, the 
number of events for the production of elements with
$Z \gtrsim 104$ is very low,
ranging from just a few to a few dozen. Consequently,
the information available for these nuclei is limited to 
their existence (where even the attribution of a neutron number is 
sometimes a matter of debate), their dominating decay channel 
(usually fission or $\alpha$ decay) and corresponding lifetime,
and in case of $\alpha$-decay also the $Q_\alpha$ value. For some odd-$A$
nuclei the branching of $\alpha$-decay into different levels 
in the daughter nucleus is observed already at this low
level of statistics.
Not only the properties of the atomic nuclei are studied for
these heavy elements, but also some of their basic chemical 
properties~\cite{Eic12a}.

Detailed spectroscopic information about their nuclear structure is 
available only for nuclei up to the lower border of the region of 
transactinides, for $Z$ values up to about $104$
\cite{Cha77a,Lei04a,Her04a,Her08a,Her11a}. These can be produced in 
much larger quantities and have longer lifetimes than most of the 
(known) heavier nuclides, such that they can be studied with a wide range 
of experimental setups. Electromagnetic transitions within and between 
rotational bands up to quite high spin are obtained from in-beam 
spectroscopy. Not only ground-state bands are studied, but there is
also an increasing body of data for bands build on $K$ 
isomers~\cite{Her06a,Cha07a}. Complementary information is 
obtained from decay spectroscopy of stopped ions. It provides
detailed information about $\gamma$ transitions from $K$ isomers,
$\alpha$-decay transition energies, 
lifetimes, branching ratios for transitions to different levels
in the daughter nuclei, and also information about 
electromagnetic transitions in the daughter nuclei. As in many
cases the daughter nuclei are $\alpha$ emitters themselves, one 
experiment usually provides data for several nuclei~\cite{Cha06a,Hes06a}. 
Finally, ion traps can be used to measure 
ground-state properties. The latter technique has been added only 
very recently to the arsenal of heavy-element research.
The first pioneering experiments delivered the masses of a few nuclides 
around \nuc{254}{No} \cite{Blo10a}.
Up to now, the quantum numbers of energy levels have been directly 
determined only for some heavy nuclei up to $Z \simeq 100$, cf.\ 
\cite{Cha77a,Her08a,Her11a} and references therein. For heavier nuclei, 
the attribution of quantum numbers has to be based on arguments 
about selection rules for transitions, which sometimes leads to 
conflicting spin assignments or even level schemes when the experiments
are performed with different set-ups and analysed by different groups.
Again, the production cross section limits spectroscopic studies to
the most favourable cases of projectile and target combinations.

The experimental efforts are closely accompanied
by theoretical studies. Mean-field models are the tools of choice for 
the description of the structure of transactinides. 
They come in two different forms, \emph{macroscopic-microscopic models} 
on the one hand, and \emph{self-consistent mean-field models} on the 
other hand. Macroscopic-microscopic models 
\cite{Nil68a,Rag78a,Nil95a,Mol94a,Sob07a,Sob11a} 
combine the "macroscopic" energy $E_{\text{mac}}$ of a droplet of nuclear 
matter and a Strutinski shell correction $E_{\text{mic}}$ that is calculated 
from the single-particle spectrum obtained from some parametrised 
single-particle potential. Both are calculated for a common shape
parametrised by a set of coefficients $\{ \alpha_i \}$, 
$E_{\text{tot}}[\{ \alpha_i \}] = E_{\text{mac}}[\{ \alpha_i \}] 
+ E_{\text{mic}} [\{ \alpha_i \}]$.
There are many variants \cite{Bar05a} that differ
in the higher-order and deformation-dependent terms of the liquid drop
and the form of the single-particle potential. A 
calculation consists in scanning the multi-dimensional energy surface 
for a given family of nuclear shapes for minima.
By contrast, the entirely microscopic self-consistent mean-field models 
\cite{Ben03a}, used in the form of HF+BCS or HFB, have as solution a set 
of single-particle wave functions 
$\{ \psi_i \}$ that minimises the binding energy 
$E_{\text{tot}}[ \{ \psi_i \} ]$ from a given effective interaction
and thereby provides local minima in the energy surface. Still, 
the surface can be mapped with the help of Lagrange multipliers. There 
are many variants that differ in the form of the effective interaction. 
The most widely used ones are the non-relativistic contact Skyrme 
\cite{Cwi96a,Cwi99a,Cwi05a,Rut97a,Bur98a,Ben99a,Kru00a,Ben01a,Dug01a,Laf01a,Ben03b,Pei05a}
and finite-range Gogny interactions \cite{Dec99a,Dec03a,Egi00a,Del06a,War12a}
as well as relativistic mean-fields (RMF)
\cite{Rut97a,Bur98a,Ben99a,Lal96a,Ben00a,Afa03a,Afa05a,Afa11a,Lit11a,Lit12a,Pra12a}
that are also called covariant density functional theory. 
The macroscopic-microscopic method can be obtained as an
approximation to self-consistent models~\cite{Bra75a}. 
In what follows, the concept of a shell correction will be used for the 
intuitive interpretation of the results from self-consistent mean-field 
calculations, even when it is not an ingredient of the model.

Both methods have their specific advantages and disadvantages that
are often complementary. The major 
technical advantages of the microscopic-macroscopic approaches are 
their numerical simplicity, that the parametrisations of the "macroscopic" 
and "microscopic" parts can be optimised separately, and that continuous 
multi-dimensional energy surfaces for fission or fusion can be easily 
constructed. 
The price to be paid is that this approach cannot describe the feedback 
of the single-particle wave functions onto the mean field, and that the
convergence of the results in terms of higher-order shape parameters
is difficult to control~\cite{Sob07a}. There is a multitude of 
parametrisations of the nuclear surface to be found in the 
literature \cite{Has88a}, some of which are better adapted to 
compact shapes, whereas others are better suited for very elongated 
shapes and the transition to situations with two centres. 
The advantages of the self-consistent models are that due to their
variational nature they automatically optimise all shape degrees of 
freedom not fixed by global symmetries and that they incorporate 
the feedback of the single-particle wave functions on the mean field. 
The latter is not only important for the description of polarisation effects 
driven by the Coulomb field, but also for the description of alignment 
effects in rotational bands and of the rearrangement of the nucleus 
when constructing $K$ isomers. Extensions of self-consistent models 
can be used to calculate correlations such as quantum fluctuations in 
shape degrees of freedom~\cite{Egi04a,Ben08LH}. In addition, their 
time-dependent variants can be used for the study of certain aspects of 
the production of heavy elements~\cite{Sim12a}. Their main disadvantage 
is that the currently used effective interactions are not yet flexible 
enough to provide the same level of accuracy for the description of
heavy nuclei as the macroscopic-microscopic models, and that the 
construction of continuous multi-dimensional deformation 
energy surfaces needed for fission studies with the help of constraints 
and Lagrange multipliers is technically challenging.

The following sections will discuss three aspects of the structure
of superheavy nuclei that are important for the extrapolation to
the the island of stability based on the available spectroscopic
information in the $Z \approx 100$ region. We will focus on 
self-consistent mean-field models. Section~\ref{sect:spherical} 
presents predictions for single-particle spectra of spherical SHE, 
whereas Section~\ref{sect:deformed} addresses deformed shell structure. 
Section~\ref{sect:correlations} will explore the role of correlations 
beyond the mean field on the structure of SHE. Finally, 
Section~\ref{sect:conclusions} summarises the discussion.

%
%
\section{Spherical shell structure}
\label{sect:spherical}

Shell structure of transactinide and superheavy nuclei is usually 
discussed in terms of the spectrum of eigenvalues  $\epsilon_\mu$ of 
the single-particle Hamiltonian $\hat{h}$
\begin{equation}
\label{eq:spe:1}
\hat{h} \, \psi_\mu (\vec{r})
= \frac{ \delta E_{\text{tot}} }
       { \delta \psi^\dagger_\mu (\vec{r}) }
= \epsilon_\mu \, \psi_\mu (\vec{r})
\end{equation}
in even-even nuclei. They reflect the mean field in the nucleus and 
provide a zeroth-order approximation to separation energies.
It has to be stressed that there 
is no unique way of defining a single-particle energy, and that most 
definitions are model-dependent. The 
shell structure of light nuclei is 
often discussed in terms of single-particle energies that are 
associated with the centroids of the strength function for nucleon 
pick-up and stripping reactions, cf.\ Ref.~\cite{Ots12a,Dug12a} 
and references therein. These cannot be expected to be 
equivalent to those from Eq.~(\ref{eq:spe:1}). 

\begin{figure}[t!]
\includegraphics[width=14.0cm]{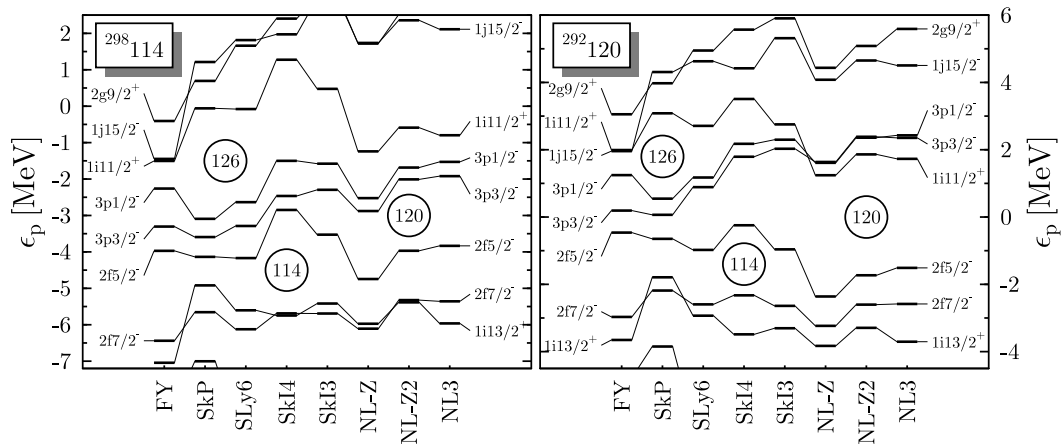}%
\vspace{-0.3cm} %
\begin{minipage}[b]{16cm}
\caption{\label{fig:spe}
Proton single-particle spectra for $^{298}_{184}114$ (left panel) and 
$^{292}_{172}120$ (right panel) obtained with
the folded-Yukawa single-particle potential used in macroscopic-microscopic
models (FY), the Skyrme interactions SkP, SLy6, SkI4 and SkI3, and the 
NL--Z, NL--Z2 and NL3 parametrisations of the RMF.
Data taken from Ref.~\cite{Ben99a}.
}
\end{minipage}
\end{figure}

Even after more than forty years of experimental and theoretical efforts,
the question where are the next closed spherical proton 
and neutron shells beyond \nuc{208}{Pb} is still open. There is no 
evidence that there are singly-magic spherical ones among the superheavy 
nuclei observed in experiment so far, and there is no consensus among 
theorists with regard to the location of the island of stability.
Concerning predictions for shell structure, the main challenges 
for theory are:
\begin{enumerate}
\item
The models have to be extrapolated far beyond the region where their
parametrisations were adjusted. The reliability of the extrapolation
of nuclear structure models has rarely been quantified so far. For
mean-field models, first steps into this direction were taken just 
recently for the extrapolation to the neutron drip line at large 
proton-to-neutron ratio \cite{Erl12b}. 
\item
The average density of single-particle levels around the Fermi 
energy increases with particle number.
As a consequence, the spacing of single-particle levels is much 
smaller than the one of lighter nuclei up to \nuc{208}{Pb}. Already a 
small relative shift  of one single-particle 
level can open up or close a gap in the spectrum. 
\item
The strong Coulomb fields try to push the protons as far apart 
from each other as possible, and thereby to polarise the nucleus. Corrections 
to the Coulomb, isovector and surface energies that play no evident
role for lighter nuclei might become crucial for SHE.
\end{enumerate}
All of the above make predictions for the shell structure of superheavy 
nuclei in the island of stability a challenging task. The current 
situation can be summarised as follows:
\begin{itemize}
\item
All widely used parametrisations of the single-particle potentials for 
macroscopic-micro\-scopic models (Nilsson, Folded-Yukawa, Woods-Saxon) 
predict \mbox{$Z=114$}, \mbox{$N=184$} as the dominant spherical shell 
closures in SHE \cite{Nil68a,Cha77a,Rag78a,Mol94a,Sob07a,Sob11a}. 
However, modifying the surface thickness of the single-particle potential
might close the \mbox{$Z=114$} gap and open up a gap at 126 
instead~\cite{And76a,Pat99a}. Note that the Woods-Saxon potential fitted by 
Chasman~\cite{Cha77a,Cha71a} contains momentum-dependent 
terms that are not considered in the work of other groups. 
\item
Self-consistent models using Skyrme, Gogny, or RMF
interactions almost always give different doubly-magic nuclei
than the standard macroscopic-microscopic models. They predict either
$Z=120$ or 126 for the next proton shell closure and $N=172$ or 184 
for the next neutron shell closure~\cite{Cwi96a,Rut97a,Ben99a}. Not 
all combinations of these magic numbers are found, however. A notable 
exception is the SkI4 Skyrme parametrisation, which predicts $Z=114$ 
and $N=184$ like the macroscopic-microscopic models~\cite{Rut97a,Ben99a}.
\end{itemize}

\begin{figure}[t!]
\begin{minipage}[b]{7.2cm}
\includegraphics[width=7.2cm]{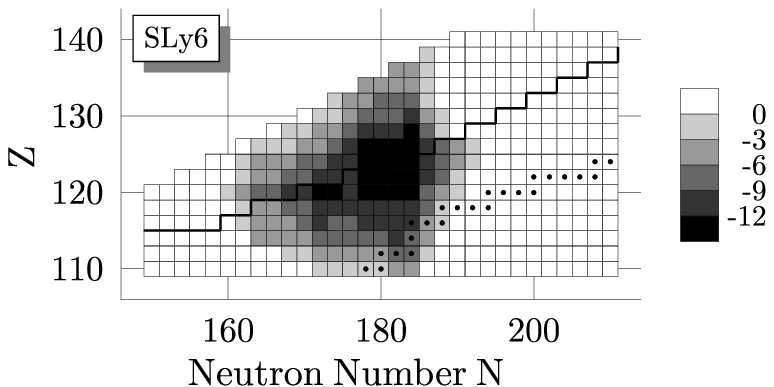}\\[2mm]
\includegraphics[width=7.2cm]{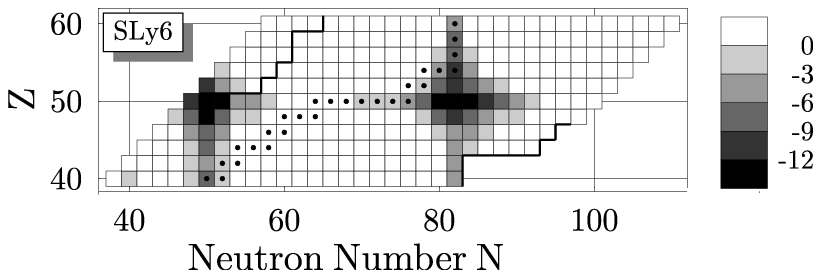}%
\vspace{-0.2cm}%
\caption{\label{fig:sc}
Total shell correction energy in MeV determined
from fully self-consistent spherical calculations with the Skyrme SLy6 
parametrisation for nuclei in the Sn region
(bottom) and the transactinide region (top). Data taken
from Ref.~\cite{Ben01a}.
}
\end{minipage}
\hspace{1.5pc}%
\begin{minipage}[b]{7.8cm}
\includegraphics[width=7.8cm]{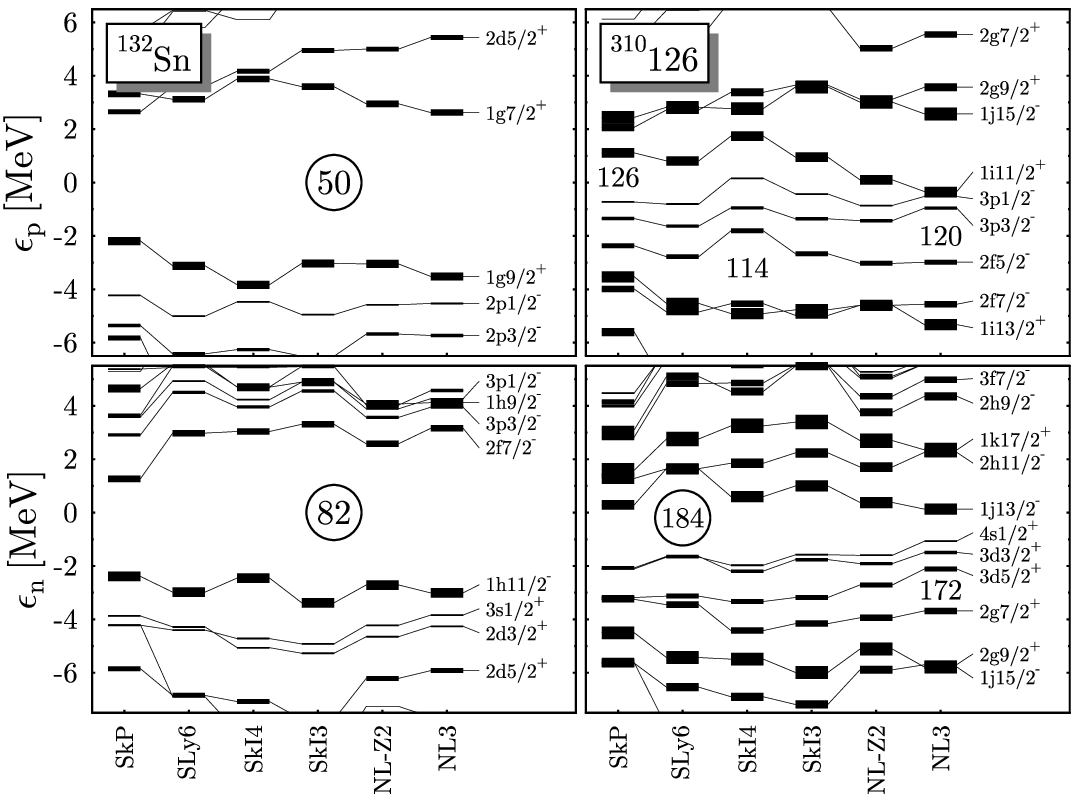}%
\vspace{-0.4cm}%
\caption{\label{fig:sc2}
Proton (top) and neutron (bottom) single-particle levels of \nuc{132}{Sn} 
(left) and $^{310}_{184}{126}$ (right) obtained with the
Skyrme (SkP-SkI3) and RMF (NL-Z2, NL3) interactions
as indicated. To illustrate the degeneracy of the 
levels, their thickness is drawn proportional to $2j+1$.
Data taken from Ref.~\cite{Ben01a}.
}
\end{minipage}
\end{figure}

\noindent
Typical results that illustrate the lack of agreement between the models
for proton single-particle states are displayed in Fig.~\ref{fig:spe}. 
The comparison between $^{298}_{184}114$ and $^{292}_{172}120$ 
illustrates also that the single-particle spectra from self-consistent 
mean-field models change rapidly with proton and neutron 
numbers in this mass region~\cite{Rut97a,Ben99a}.
For example, the gap at $Z=120$ is much larger when the proton number 
takes this value than for the $Z=114$ isotope. A similar effect is 
found for the neutron shell closure at $N=172$~\cite{Ben99a}. 
As can bee seen in Fig.~\ref{fig:spe}, predictions for the gaps in 
the single-particle spectra rarely exceed 2~MeV. In fact, it has been
pointed out that the actual gaps themselves are no longer the decisive 
criterion for magicity in SHE~\cite{Ben01a}. This is illustrated in 
Fig.~\ref{fig:sc} that shows the shell correction determined from
fully self-consistent mean-field calculations with the Skyrme interaction
SLy6 for the tin region and for superheavy nuclei. For nuclei in the 
Sn region, the combined shell corrections for protons and neutrons are
localised at the magic numbers $Z=50$, $N=50$ and $N=82$. They take their
largest values for the doubly-magic nuclei \nuc{100}{Sn} and \nuc{132}{Sn}, 
and fall of quickly when going away from them. The situation is very 
different for superheavy nuclei: here, the shell correction takes large 
values in a large region that ranges from $N \approx 164$ to $N \approx 184$ 
and $Z \approx 114$ to $Z \approx 126$, and that encompasses all possible
predictions for doubly-magic superheavy spherical nuclei. 
This finding is fairly independent of the chosen effective interaction 
\cite{Ben01a}. It can be understood when recalling that the spherical 
single-particle levels have a $(2j+1)$-fold degeneracy. The levels 
near the Fermi energy displayed in Fig.~\ref{fig:spe} range from 
$j=1/2$ up to $15/2$, and their degeneracy from 2 to 16. Its role
can be better seen in the single-particle spectra displayed
in Fig.~\ref{fig:sc2}, where the width of the levels is drawn 
proportional to their degeneracy. For spherical superheavy nuclei, 
the low-$j$ orbits of protons and neutrons are bunched amidst a
large number of high-$j$ levels. Due to their large degeneracy, 
it are the latter that determine the average level density.
The low-$j$ levels can be shifted around opening up or closing the
gaps between them without making a large difference for the level 
density and the size of the shell correction, even when 
the corresponding single-particle spectra look very different.
The low-$j$ are filled for proton numbers between 114 and 126 and neutron 
numbers between 172 and 184, which is exactly the zone where the 
combined shell correction is large. The appearance of bunched low-$j$ 
levels surrounded by 
high-$j$ levels follows trivially from the sequence of single-particle 
levels in a finite potential well and the spin-orbit splitting in
nuclei that is large enough to push an intruder into the major shell
below.

\begin{figure}[t!]
\begin{minipage}[b]{7.2cm}
\includegraphics[width=7.2cm]{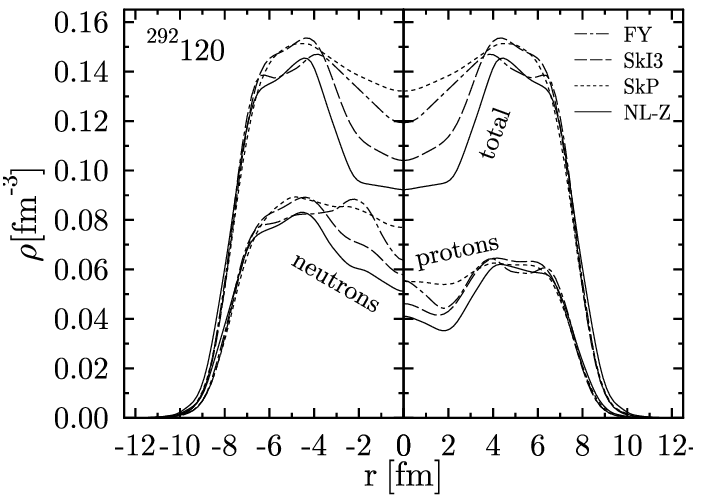}%
\vspace{-0.4cm}%
\caption{\label{fig:dens}
Radial profile of the neutron (left), proton (right) and total density 
of the spherical configuration of $^{292}_{172}120$, obtained with the 
Skyrme interactions SkI3 and SkP, the RMF Lagrangian
NL-Z and the folded-Yukawa (FY) single-particle potential widely used
in macroscopic-microscopic calculations. Data taken from Ref.~\cite{Ben99a}.
}
\end{minipage}
\hspace{1.5pc}%
\begin{minipage}[b]{7.8cm}
\includegraphics[width=7.8cm]{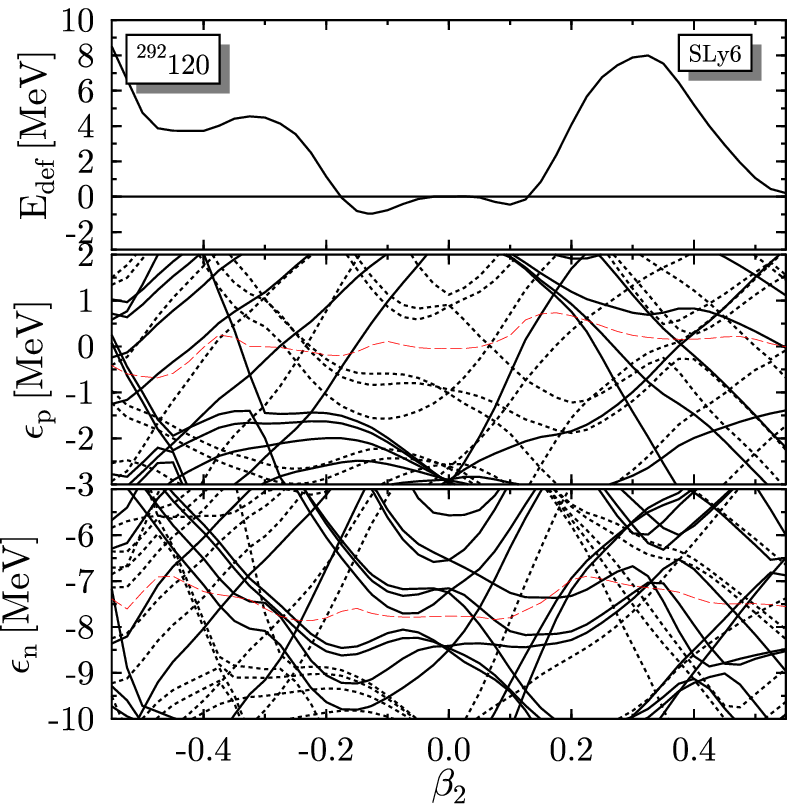}%
\vspace{-0.4cm}%
\caption{\label{fig:172:120:pes}
Axial deformation energy surface (top) and Nilsson diagrams of
protons (middle) and neutrons (bottom) as a function of
quadrupole deformation $\beta_2$, Eq.~(\ref{eq:beta2}), 
as obtained with the Skyrme
interaction SLy6 for $^{292}_{172}120$. In the Nilsson diagrams,
solid (dotted) lines denote levels of positive (negative) parity.
The Fermi energies of protons and neutrons are indicated with 
dashed red lines.
}
\end{minipage}
\end{figure}

The appearance of the magic numbers $Z=120$ and $N=172$ in 
self-consistent models is the consequence of a polarisation of the 
density~\cite{Ben99a,Pei05a,Dec99a,Dec03a,Afa05a}. 
The low-$\ell$ $s$, $p$, and $d$ orbits 
that were found to be crucial for the extra binding of spherical 
superheavy nuclei in the discussion above have as a common feature that 
their radial wave functions are localised well inside the nucleus. In 
general, only low-$\ell$ levels contribute to the density at small 
radii. Ignoring pairing and other correlations, $^{292}_{172}120$ is the 
last nucleus for which they are not filled. By contrast, the occupied
high-$\ell$ levels in the last major shell are all peaked on the 
the nuclear surface. Combining both, the net effect is a very pronounced 
central depression of both the proton and neutron densities, cf.\ 
Fig.~\ref{fig:dens}, that has been characterised as a "semi-bubble" 
density profile~\cite{Dec99a,Dec03a}. In self-consistent models the 
density distribution feeds back onto the mean fields, in particular onto
the radial profile of the spin-orbit potential~\cite{Ben99a}. For 
specific orbits such as the proton $3p$ and neutron $3d$ levels, the 
spin-orbit splitting is reduced to almost zero, cf.\ Fig.~\ref{fig:spe},
which opens up the $Z=120$ and $N=172$ gaps. When filling the low-$j$ 
orbits one goes back to a regular density distribution and the gaps at 
$Z=120$ and $N=172$ disappear~\cite{Ben99a}. Similar, but weaker, 
effects can be seen 
in lighter nuclei. For those it is usually just the non-occupation of 
one proton or neutron low-$j$ level that causes a central depression 
of the density. The new feature of \nuc{292}{120} is the much larger
number of single-particle states and nucleons involved that makes the effect 
more pronounced. The size of this polarisation effect is correlated 
to the effective mass. It is the largest for small effective mass 
(NL-Z and SkI3 with $m_0^*/m \approx 0.6$) and less pronounced 
for large effective mass (SkP, $m_0^*/m = 1$). Interestingly, 
the density distribution from the folded-Yukawa (FY) single-particle 
potential  frequently used in macroscopic-microscopic methods also 
exhibits a central depression, cf.\ Fig.~\ref{fig:dens}. However, 
in the absence of self-consistency it neither feeds back onto the 
shell structure, nor onto the macroscopic energy.

\begin{figure}[t!]
\begin{minipage}[b]{16.0cm}
\includegraphics[width=14.0cm]{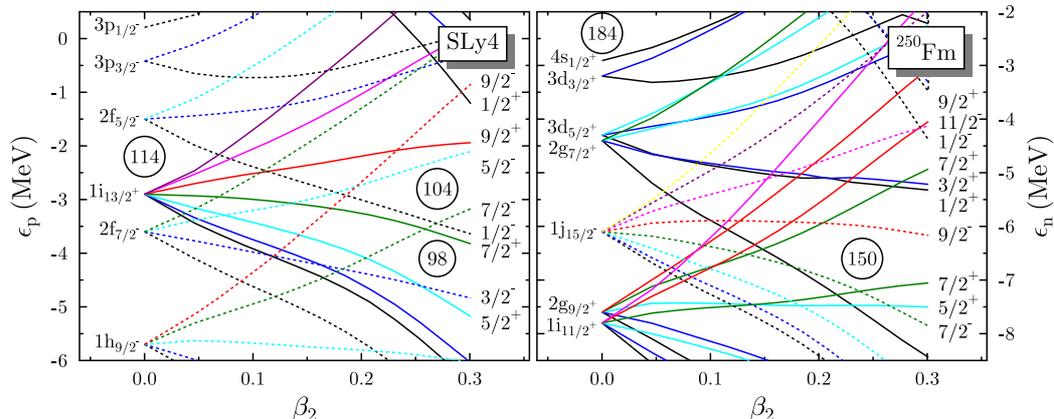}%
\vspace{-0.4cm}%
\caption{\label{fig:nilsson}
Nilsson diagrams of proton (left) and neutron (right) single-particle 
levels of $^{250}_{150}\text{Fm}_{100}$ as obtained with the SLy4 
Skyrme parametrisation \cite{Cha98a} as a function of
quadrupole deformation going from spherical shape to the 
prolate ground-state deformation. Labels on the left denote spherical
$j$ shells, labels on the right indicate $j_z$ and parity. Levels of
positive (negative) parity are drawn as solid (dotted) lines. The
colours indicate different mean values of $j_z$.
}
\end{minipage}
\end{figure}

Large gaps in the single-particle spectrum do not necessarily imply
a spherical ground state.
This is illustrated for $^{292}_{172}{120}$ in Fig.~\ref{fig:172:120:pes}.
The deformation energy surface is quite soft around the spherical 
point ($\beta_2 = 0$), with an oblate minimum at a $\beta_2$ value of $-0.13$. 
Its softness is again related to the presence of low-$j$
levels around the Fermi energy for protons and neutrons. The splitting 
of their sub-states with deformation is rather weak and
many of the levels on both sides of the Fermi surface repel each other 
such that the $Z=120$ and $N=172$ gaps persist far to the oblate side.
The prediction whether \nuc{292}{120} has a spherical or a deformed ground 
state, however, is very sensitive to details of the effective interaction. 
The Gogny force~\cite{Dec99a,Dec03a,War12a} and most Skyrme 
interactions~\cite{Pei05a,Ben98a} give a similar result as SLy6, whereas 
a few other Skyrme interactions~\cite{Pei05a} and most relativistic mean-field 
models give a spherical minimum~\cite{Ben98a}.

There is a large number of non-equivalent definitions of quadrupole 
deformations to be found in the literature, which is a source of 
confusion and misunderstandings 
when comparing results of different groups. Throughout this
article we will use the dimensionless quadrupole deformation $\beta_2$
that is obtained from the mean value of the quadrupole moment taking
out its trivial scaling with proton and neutron number
\begin{equation}
\label{eq:beta2}
\beta_2
\equiv \frac{4\pi}{3R^2 A} \, \langle Q_{2 0} \rangle
= \frac{4\pi}{3R^2 A} \, \sqrt{\frac{5}{16\pi}} \,
  \langle 2 z^2 - x^2 - y^2 \rangle
\, ,
\end{equation}
where $R = 1.2 \, A^{1/3}$ fm. 
Results from macroscopic-microscopic methods are usually shown 
in terms of the multipole expansion parameters of the nuclear 
surface, for which there exist several non-equivalent conventions
and which are not equivalent to the mean values of the 
multipole moments \cite{Has88a}. Results from self-consistent
models have also sometimes been presented in terms of shape expansion 
parameters estimated from the multipole moments~\cite{Cwi96a,Cwi05a}.

%
%
\section{Deformed shell structure}
\label{sect:deformed}

\begin{figure}[t!]
\begin{minipage}[b]{16.0cm}
\includegraphics[width=5.2cm]{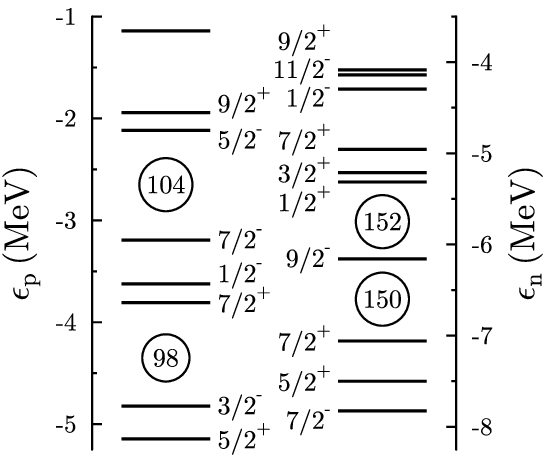}\hspace{0.2cm}%
\includegraphics[width=5.2cm]{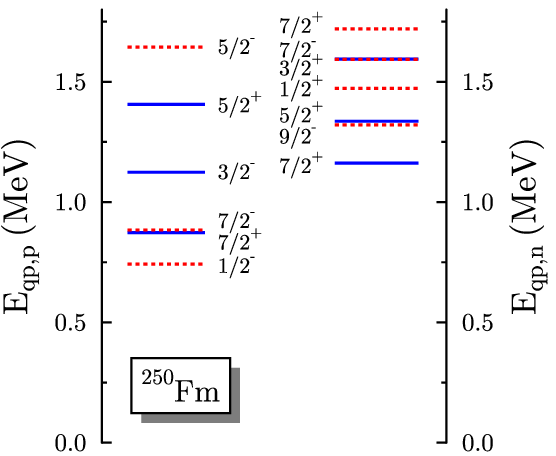}\hspace{0.2cm}%
\includegraphics[width=5.2cm]{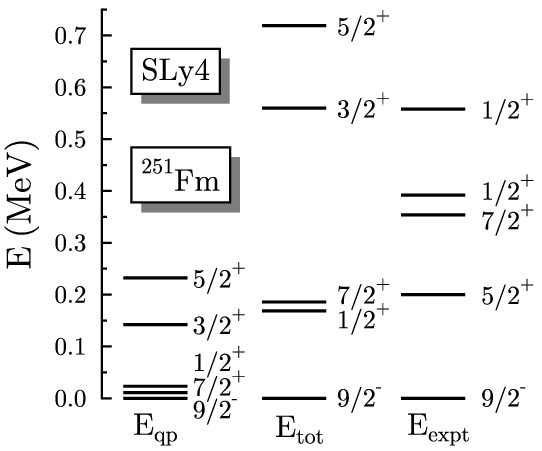}%
\vspace{-0.3cm}%
\caption{\label{fig:Fm250:Eqp}
Spectrum of eigenvalues of the mean-field (left, Eq.~(\ref{eq:spe:1})) 
and HFB Hamiltonian (middle, Eq.~\ref{eq:eqp:1})) for protons and neutrons 
for the deformed ground state of \nuc{250}{Fm} as obtained with the SLy4 
Skyrme parametrisation. Solid blue (dashed red) lines indicate eigenvalues 
of the HFB Hamiltonian corresponding to single-particle levels below 
(above) the Fermi energy. The right panel compares the spectrum of
eigenvalues $E_{\text{qp}}$ of the HFB Hamiltonian for neutrons obtained 
for a HFB vacuum constrained to \mbox{$Z = 100$}, \mbox{$N = 151$} 
and the spectrum of the energies from self-consistently 
calculated blocked HFB states $E_{\text{tot}}$ in \nuc{251}{Fm}
with experimental data for band heads taken from \cite{Her08a}.
All energies in the right panel are normalised to the ground state.
}
\end{minipage}
\end{figure}

We have seen above that there exists a multitude of conflicting 
predictions for the shell structure of spherical superheavy nuclei that 
constitute the "island of stability". As of today, the 
question about the location of the next shell closures cannot be
answered by theory alone, and there is an ongoing effort to use the
available spectroscopic data for well-deformed transactinides in 
the $Z \approx 102$ region. The basic idea can be illustrated by
the Nilsson diagram shown in Fig.~\ref{fig:nilsson}, which covers
deformations ranging from spherical shape to the prolate deformed ground
state. At the ground-state deformation, the single-particle 
levels can be associated with the observed band heads of 
rotational bands in odd-$A$ nuclei. Measuring the energy differences 
between these band heads in the same nucleus and having a mean-field
model that describes their relative positions, one can then estimate
empirical spherical single-particle energies by tracing the
Nilsson diagram back to spherical shape.
The single-particle energies as they appear in the Nilsson
diagram, however, are not observable. The differences 
between the energies of the various band heads measured
differ from the single-particle energies in several respects.
First, pairing correlations introduce a correction to the relative
energies. Second, the unpaired nucleon polarises the mean fields. 
In particular, due to the lifting of Kramers degeneracy in an 
odd-$A$ nucleus, there is no useful way of representing 
single-particle energies as defined in Eq.~(\ref{eq:spe:1}) for 
the purpose of analysing the shell structure. 
Third, each configuration in an odd-$A$ nucleus has in general 
its own ground-state deformation. The effects of self-consistency 
when calculating these band heads are illustrated 
in Fig.~\ref{fig:Fm250:Eqp}. The left panel shows the single-particle
spectra obtained for the ground state of \nuc{250}{Fm}, which correspond 
to the end of the Nilsson diagrams in Fig.~\ref{fig:nilsson}. Pairing 
correlations modify their relative energies. These are 
incorporated in the spectrum of eigenvalues $E_{\text{qp}}$
of the HFB Hamiltonian 
\begin{equation}
\label{eq:eqp:1}
\fourmat{h        }{\Delta} 
        {-\Delta^*}{-h^*  } \,
\twospinor{U}{V}
= E_{\text{qp}} \, \twospinor{U}{V}
\, ,
\end{equation}
often called quasi-particle energies and shown in the middle panel of 
Fig.~\ref{fig:Fm250:Eqp}. When neglecting off-diagonal matrix
elements in the HFB equation~(\ref{eq:eqp:1}), which leads to the simpler 
HF+BCS scheme, the quasi-particle energy of a single-particle 
level represents its distance from the Fermi energy $\lambda$ and 
the size of its pairing gap $\Delta_{\mu \bar\mu}$ through the relation
\mbox{$E_{\text{qp}} \approx
\sqrt{( \epsilon_\mu - \lambda)^2 + \Delta_{\mu \bar\mu}^2}$}. For
a realistic pairing interaction, the values of the gaps 
$\Delta_{\mu \bar\mu}$ strongly depend on the single-particle level.
In full HFB, the coupling of the quasi-particle states through non-diagonal 
matrix elements slightly modifies the spectrum. 
The right panel of Fig.~\ref{fig:Fm250:Eqp} compares two different
calculations of the band heads in \nuc{251}{Fm} with data. One is a
perturbative calculation of these states from the spectrum of quasi-particle 
energies $E_{\text{qp}}$ obtained from Eq.~(\ref{eq:eqp:1}). The difference
to the neutron spectrum in the middle panel of Fig.~\ref{fig:Fm250:Eqp} is
that the HFB vacuum is now calculated for particle numbers \mbox{$Z = 100$}, 
\mbox{$N = 151$}. The readjustment of the Fermi energy for neutrons
visibly rearranges the spectrum. In particular, the lowest quasi-particle 
state is now a $9/2^-$, in agreement with data. However, compared to 
experiment the spectrum of the $E_{\text{qp}}$ is too compressed.
Such perturbative calculation of low-lying states presents at best a crude
approximation to the energies from full self-consistent calculations of 
the blocked HFB state for an odd-$A$ nucleus \cite{Dug02a,Ber09a}, 
which is represented
as $E_{\text{tot}}$ in the right panel. In this case, the energy of each
configuration results from a separate calculation of the quasi-particle
excitation performed on an optimised vacuum.
There are many rearrangement effects, such as a reduction 
of pairing when blocking a state, and changes in the mean fields and 
deformations, which are also all different for each blocked configuration.
The net effect is that the spectrum of the $E_{\text{tot}}$ is markedly
different: the level sequence changes, and the spectrum is more
spread out. The calculated level sequence, however, does not agree
with data. In particular, the $5/2^+$ level should be lower and
the $1/2^+$ level higher. This can be achieved by closing the 
\mbox{$N=150$} gap in the single-particle spectrum and making the 
\mbox{$N=152$} gap larger. The same modification of the single-particle
spectrum is suggested by the analysis of adjacent 
odd-$A$ nuclei and by two-quasi-particle states observed 
as $K$ isomers in even-even nuclei \cite{Her08a}.
The presence of a deformed \mbox{$N=150$} gap not seen in experiment
in connection with a
too small \mbox{$N=152$} gap is a common problem of all
self-consistent mean-field models using Skyrme \cite{Bur98a,Ben03b},
Gogny \cite{Del06a} and relativistic mean-field interactions 
\cite{Bur98a,Afa03a,Afa11a}.
Similar problems are found for the single-particle spectra of protons,
where a gap should be at \mbox{$Z = 100$}, not \mbox{$Z = 98$}
and \mbox{$Z = 104$} \cite{Cha06a,Bur98a,Ben03b,Del06a,Afa03a,Afa11a}.
The deformed gaps are fairly reproduced by the single-particle potentials
used in macroscopic-microscopic calculations, which explains their
superiority for describing details of the spectroscopic data in the
\mbox{$A \approx 250$} region.

\begin{figure}[t!]
\begin{minipage}[b]{14.0cm}
\includegraphics[width=7.0cm]{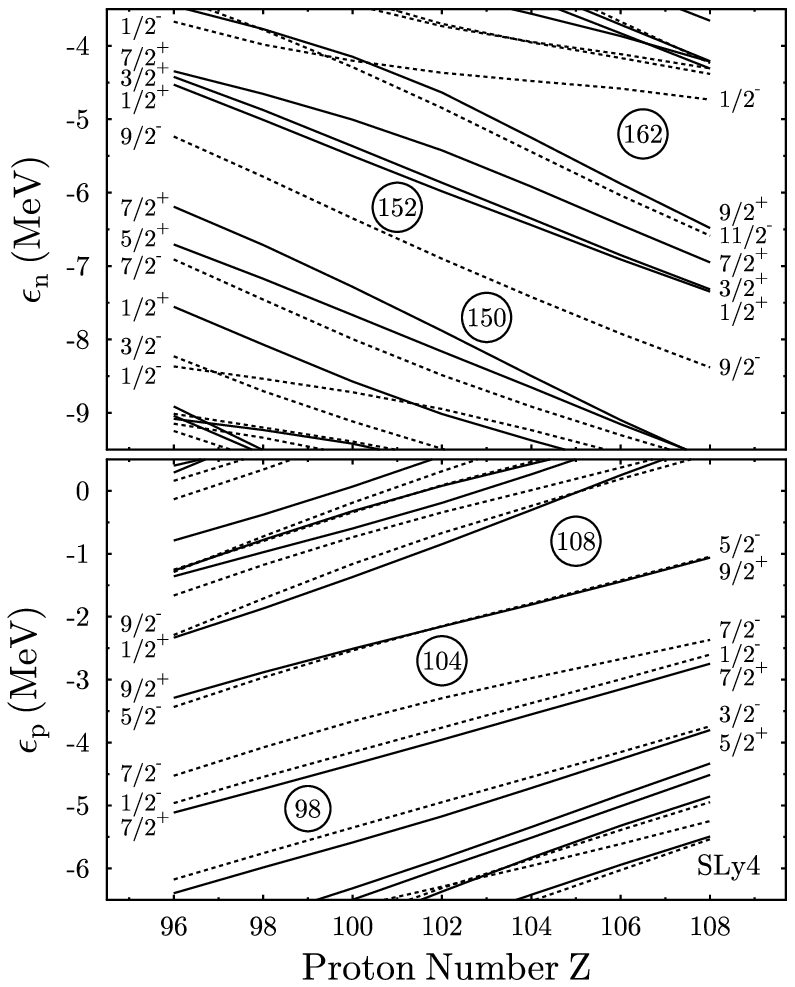} 
\includegraphics[width=7.0cm]{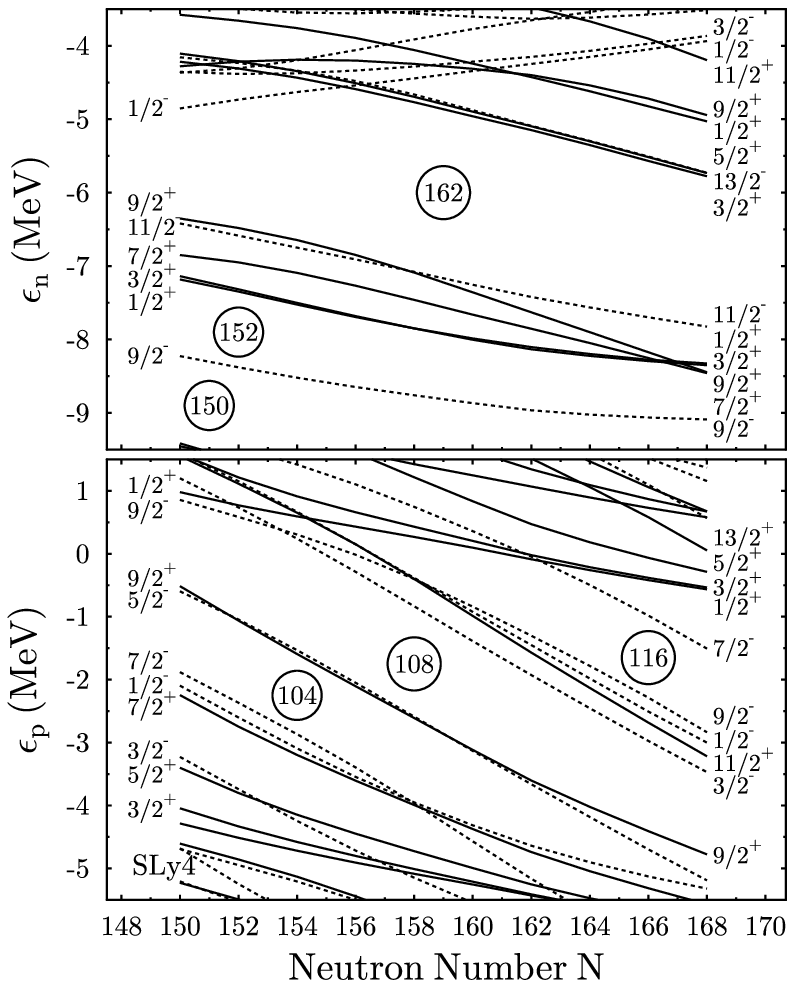}%
\vspace{-0.4cm}
\caption{\label{fig:spe:def:min}
Single-particle spectra of neutrons (top panel) and protons (bottom) 
for the deformed ground states of $N=152$ isotones
as a function of proton number (left) and of Hs ($Z=108$) isotopes as a 
function of neutron number (right) calculated with the SLy4 
parametrisation of Skyrme's interaction. 
}
\end{minipage}
\end{figure}

\begin{figure}[t!]
\includegraphics[width=6.0cm]{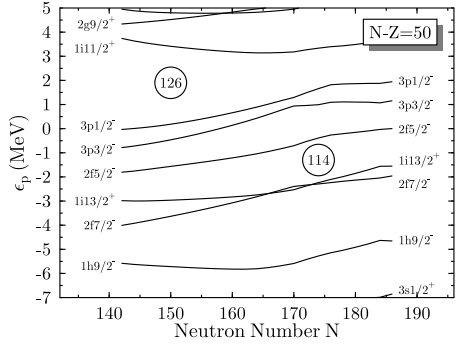} \hspace{2pc}%
\begin{minipage}[b]{8.0cm}
\caption{\label{fig:extra:sphere}
Evolution of \emph{spherical} proton shell structure as predicted
by the Skyrme SLy4 parametrisation when going from heavy actinides 
to superheavy nuclei along the $\alpha$-decay chain with \mbox{$N-Z=50$}. 
For most nuclei shown, the spherical configuration corresponds to a local
\emph{maximum} of the energy surface, not a minimum. The spherical 
configuration of \nuc{250}{Fm}, i.e.\ $\beta_2 = 0$ in the left panel 
of Fig.~\ref{fig:nilsson}, corresponds to \mbox{$N=150$}.
}
\end{minipage}
\end{figure}

In any event, the single-particle spectra, in particular in the form
of Nilsson diagrams, are a more intuitive tool for the interpretation
of shell structure than the spectra from self-consistent 
blocked states. We will use them in what follows to 
illustrate two further difficulties of establishing the link
between deformed and spherical shell structures.
The first of these difficulties is that the analysis cannot 
be done globally using the band heads of nuclei throughout
the region of heavy nuclei. Instead, 
it has to be done locally, using only data from one or perhaps a 
few adjacent nuclei. The reason is that there are substantial 
changes in the single-particle spectra of the deformed ground 
states when varying $N$ and $Z$, cf.\ Fig.~\ref{fig:beta2}.
The nucleons rearrange themselves for the maximum gain in binding 
energy that is possible for a given number of particles. In 
particular, some of the deformed shell closures such as \mbox{$Z=116$}
and \mbox{$N=162$} open up only for specific combinations of $N$ and $Z$. 
The rearrangement of shells is accompanied by slight changes in all 
multipole moments and the density profile.
The second difficulty is that also the spherical shell structure
changes rapidly for heavy nuclei. This is illustrated by 
Fig.~\ref{fig:extra:sphere}. Assuming for the moment that the effective 
interaction used would correctly describe the shell structure of nuclei 
around $^{250}_{150}\text{Fm}_{100}$, one would conjecture 
from the Nilsson diagram in Fig.~\ref{fig:nilsson} that there is a strong 
spherical \mbox{$Z \! = \! 126$} shell closure. The \mbox{$Z \! = \! 126$} 
gap, however, shrinks substantially when going to heavier nuclei, which
is the effect already discussed in Fig.~\ref{fig:spe}. 
In general, what is deduced about the magic numbers in the spherical 
single-particle spectrum in a non-magic nucleus might not be valid 
anymore for nuclei that actually have this magic number. 
Even when the SLy4 parametrisation does not describe all details
of deformed shell structure in the $Z \! \approx \! 100$ region, 
the rearrangement effects are generic to all self-consistent 
mean-field models.

The nucleon-number dependence of spherical shells as predicted by 
mean-field interactions is much stronger for SHE than for light elements,
where the single-particle energies evolve much more slowly. In light
nuclei, the observed strong quenching of various signatures of shell 
closures is only obtained when going beyond the self-consistent mean 
field \cite{Fle04a,Ben06a,Ben08a,Del10a}.

\begin{figure}[b!]
\begin{minipage}[b]{16.0cm}
\includegraphics[width=5.0cm]{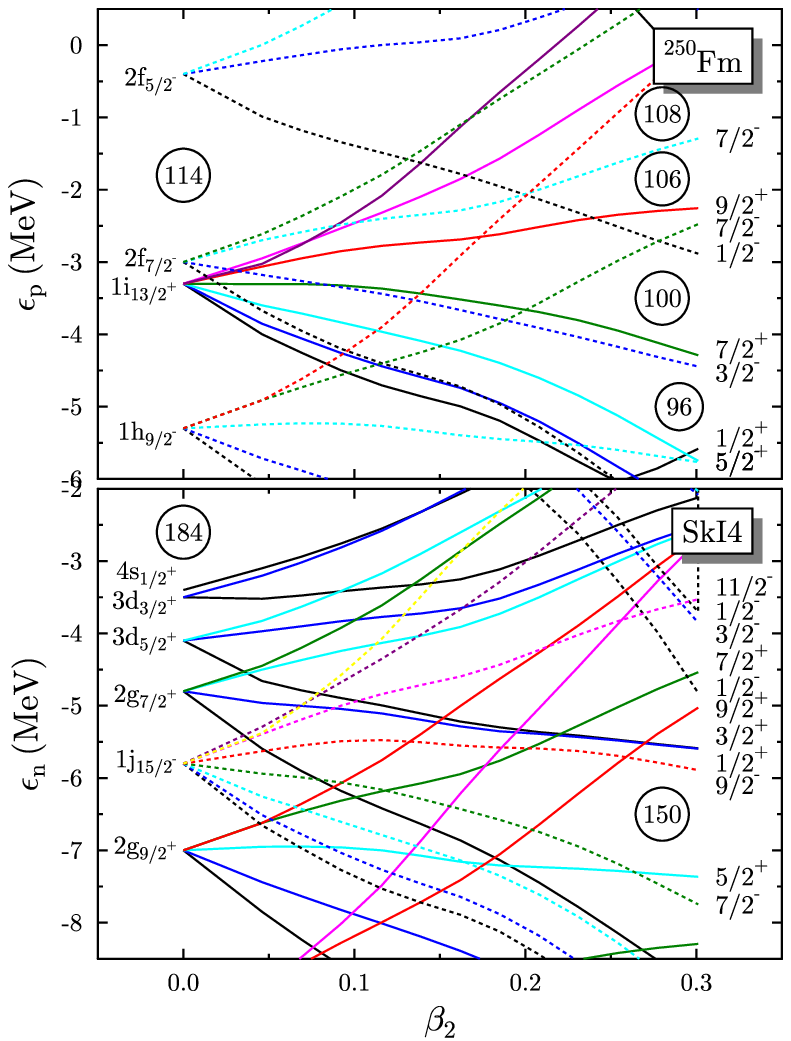}%
\includegraphics[width=5.0cm]{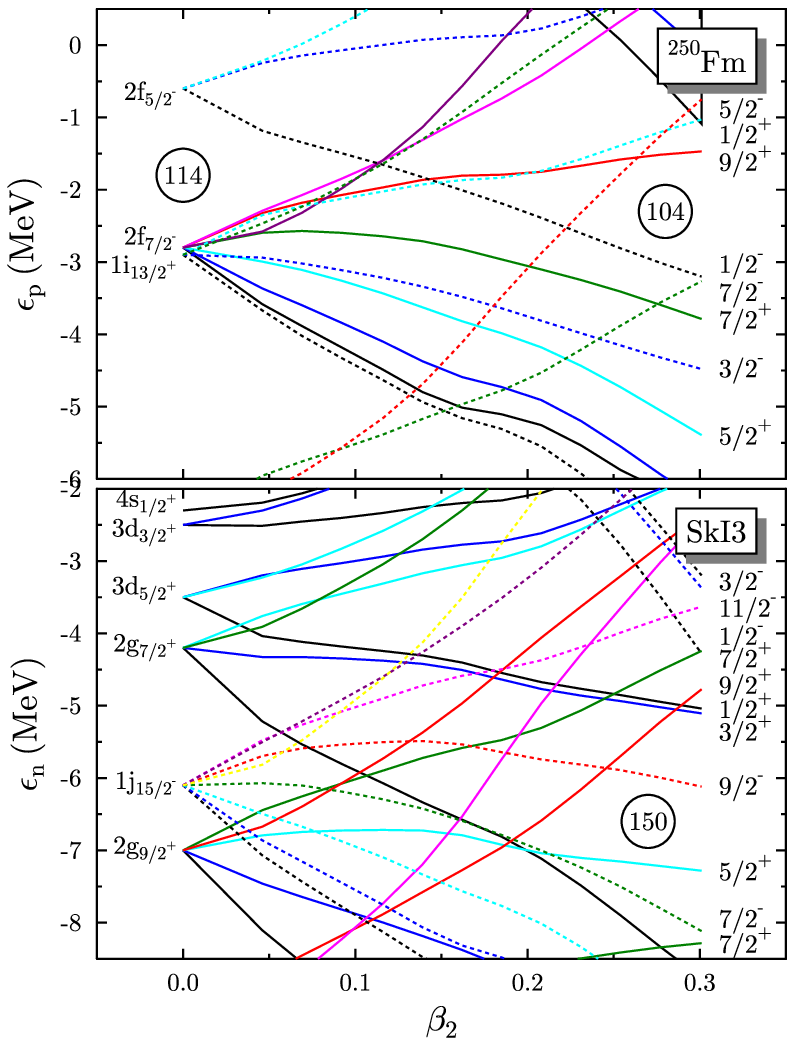}%
\includegraphics[width=5.0cm]{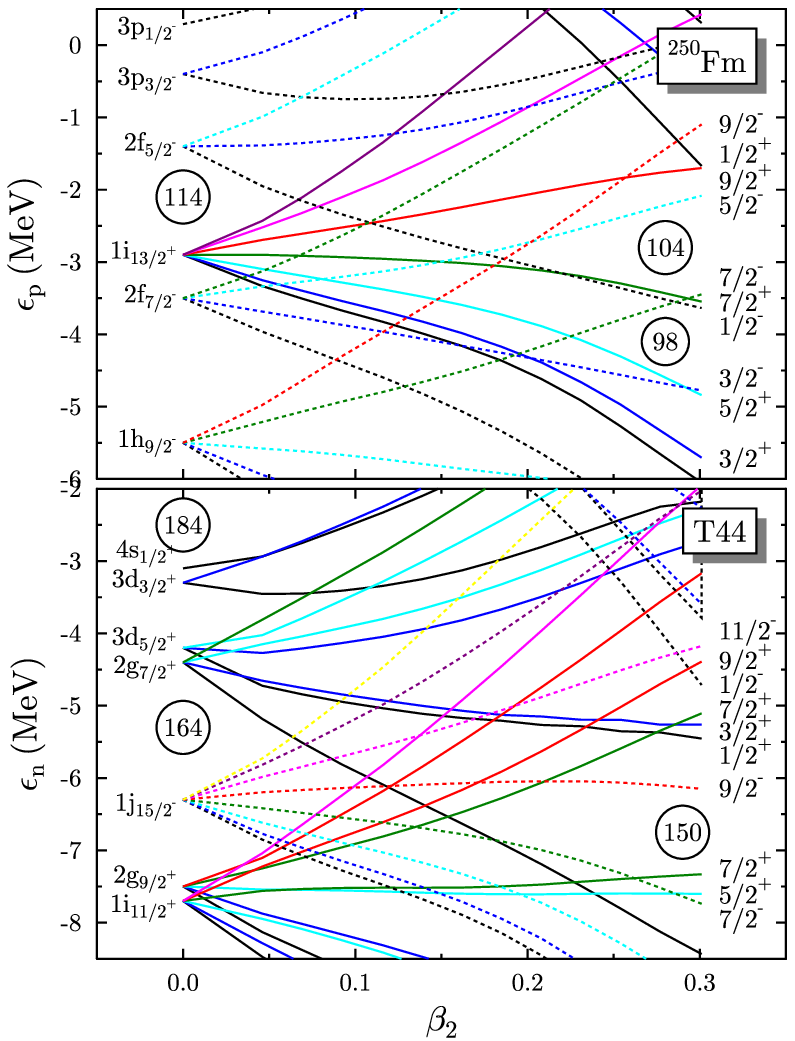}%
\vspace{-0.4cm}%
\caption{\label{fig:def:gs:2}
Parametrisation dependence of the Nilsson diagram of \nuc{250}{Fm} for
the example of three parametrisations of Skyrme's interaction. 
Left to right: SkI4 and SkI3 with two different non-standard choices 
of the isospin dependence of the spin-orbit field~\cite{Rei95a} and 
T44, which adds an explicit tensor force to a standard spin-orbit
interaction~\cite{Les07a}.
}
\end{minipage}
\end{figure}

The deformed single-particle spectra are as sensitive to details
of the effective interaction as are the spherical ones. Examples
are given in Fig.~\ref{fig:def:gs:2} for Skyrme parametrisations with
modified spin-orbit interactions. There are considerable changes 
relative to the Nilsson diagrams obtained with SLy4, cf.\
Fig.~\ref{fig:nilsson}.
However, for none of these parametrisations the deformed \mbox{$N=152$} 
gap opens up. One improvement is found for SkI4, which gives a deformed 
\mbox{$Z = 100$} gaps like the macroscopic-microscopic models. This improves
one- and two-quasi-particle spectra for proton excitations in the 
\mbox{$A \approx 250$} region \cite{Ben12SHE}. However, it comes at 
the price of closing completely the \mbox{$N=152$} gap, which deteriorates
the spectra for neutron excitations \cite{Ben12SHE}.

\begin{figure}[t!]
\begin{minipage}[b]{7.8cm}
\centerline{\includegraphics[width=6.5cm]{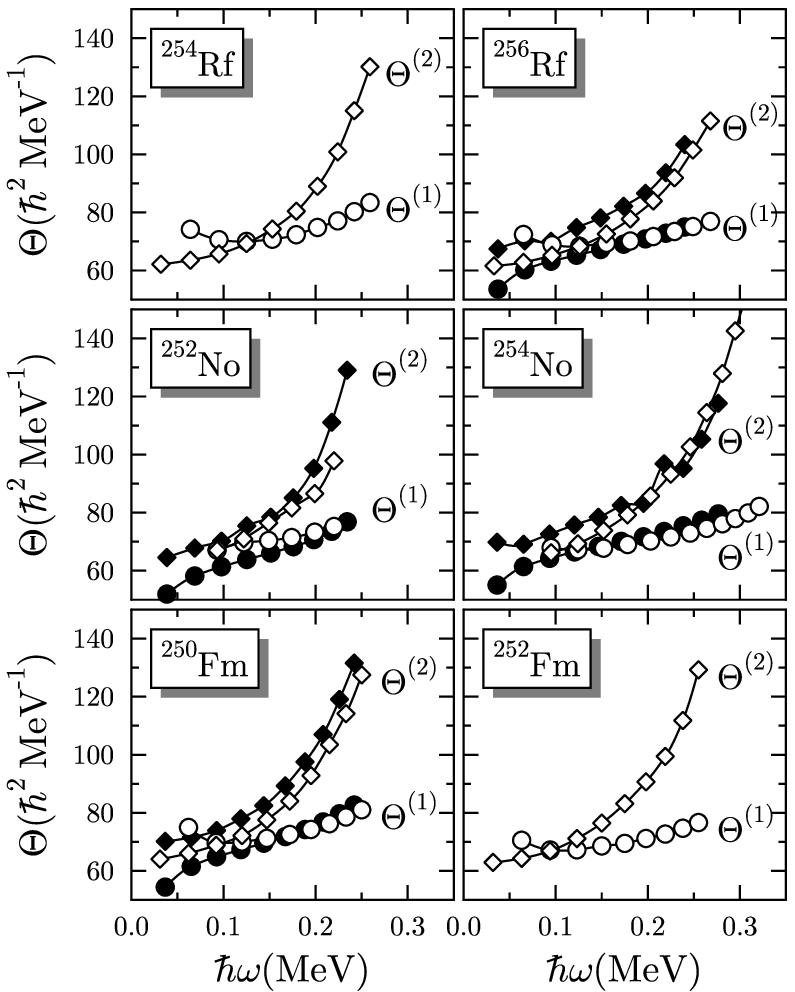}}%
\vspace{-0.4cm}%
\caption{\label{fig:rot:mom}
Kinematic ($\Theta^{(1)}$, circles) and dynamical ($\Theta^{(2)}$, diamonds) 
moments of inertia deduced from experiment (filled markers) and cranked HFB
calculations with the SLy4 Skyrme parametrisation (open markers). 
Calculated values 
taken from Ref.~\cite{Ben03b}, experimental values from \cite{Her08a,Her11a}
(\nuc{250}{Fm}, \nuc{252,254}{No}) and \cite{Gre12a,Rub12a} (\nuc{256}{Rf}).
}
\end{minipage} \hspace{2pc}%
\begin{minipage}[b]{7.0cm}
\includegraphics[width=6.5cm]{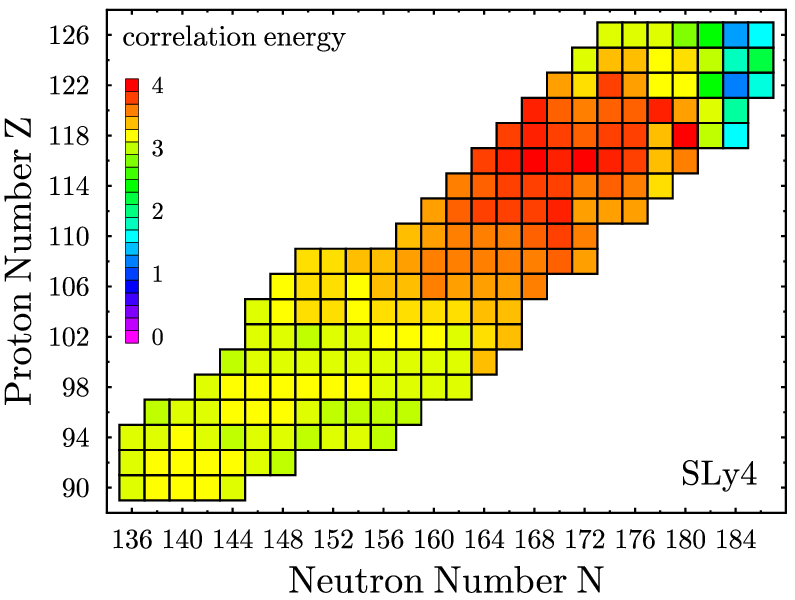}%
\vspace{-0.4cm}%
\caption{\label{fig:corr}
Quadrupole correlation energy in MeV from projection on angular 
momentum \mbox{$J=0$} and mixing of axial configurations of 
different quadrupole deformation \cite{Ben12SHE}. The correlation 
energy is defined as the difference between the energy of the 
particle-number projected mean-field ground state and the 
energy of the particle-number and angular-momentum projected
GCM ground state, cf.~\cite{Ben06a}.
Calculations were done with the SLy4 Skyrme
parametrisation. 
}
\end{minipage}
\end{figure}

The proof of large ground-state deformations in this region of the nuclear 
chart has been delivered by the observation of highly collective rotational 
bands, first for \nuc{254}{No} \cite{Rei99a}, and then for some of its 
even-even and odd-$A$ neighbours \cite{Her08a,Her11a}. The observation of 
these bands 
triggered numerous cranked HFB calculations, where the mean-field equations 
are solved with an auxiliary condition on the mean value of angular momentum 
\cite{Dug01a,Laf01a,Ben03b,Egi00a,Afa03a}. As an example, 
Fig.~\ref{fig:rot:mom} compares calculations using the SLy4 Skyrme
parametrisation with data for some even-even nuclei. The figure 
displays the kinematical (first) moment of inertia 
$\Theta^{(1)} \equiv J/\omega$ and the dynamical (second) 
moment of inertia $\Theta^{(2)} \equiv dJ / d\omega 
= \Theta^{(1)} + \omega \; d \Theta^{(1)} / d \omega$ \cite{Her08a,Dud92a} 
as a function of the rotational frequency $\omega$. The dynamical 
moment of inertia $\Theta^{(2)}$ is a sensitive probe for the internal 
rearrangement of the nucleons with increasing angular momentum $J$.
The calculations agree reasonably
well with the data, although there are disagreements in detail, for example 
the slightly different up-bend of $\Theta^{(2)}$ seen for \nuc{252}{No} and 
\nuc{254}{No}, which can be attributed to the deficiencies in 
the single-particle spectrum discussed above.

%
%
\section{Correlations beyond the mean field}
\label{sect:correlations}

\begin{figure}[t!]
\begin{minipage}[b]{7.5cm}
\includegraphics[width=6.8cm]{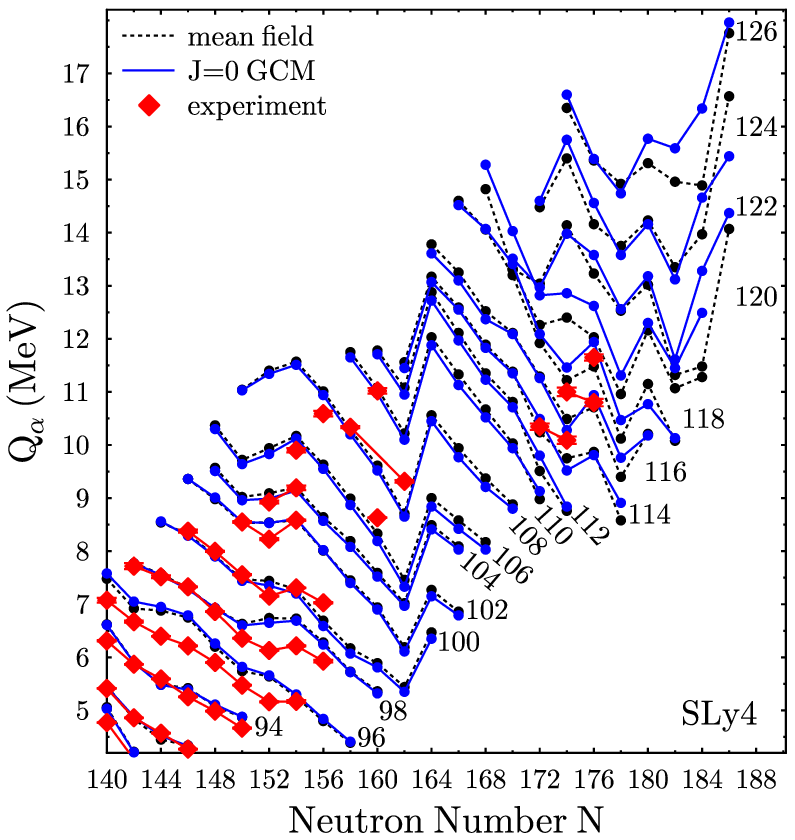}%
\vspace{-0.4cm}%
\caption{\label{fig:qa}
$Q_\alpha$ values of even-even nuclei from self-consistent 
mean-field and beyond-mean-field calculations  \cite{Ben12SHE}
with the Skyrme interaction 
SLy4 compared with data where available \cite{SHEnet}. Lines connect 
values for isotopic chains as indicated by the labels.
}
\end{minipage} \hspace{2pc}%
\begin{minipage}[b]{7.5cm}
\includegraphics[width=6.5cm]{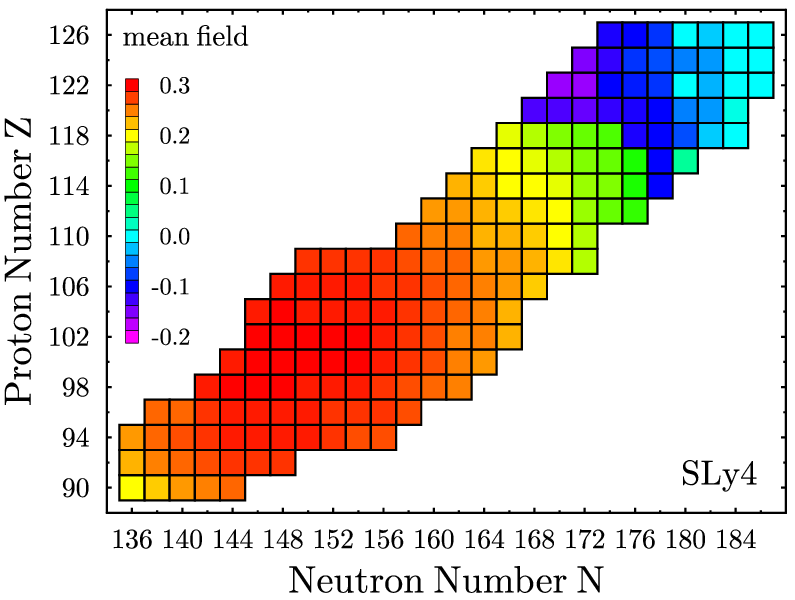}%
\vspace{-0.2cm}%
\caption{\label{fig:beta2}
Dimensionless mass quadrupole moment $\beta_2$ (\ref{eq:beta2})
of the ground state of heavy nuclei as obtained with the Skyrme 
interaction SLy4. Positive values denote prolate, negative values oblate
deformations, whereas $\beta_2 = 0$ indicates spherical shapes. Data
taken from \cite{Ben12SHE}.
}
\end{minipage}
\end{figure}

The mean-field approximation is best justified for heavy nuclei
whose ground state corresponds to a deep minimum in a stiff deformation 
energy surface. This is indeed the case for many of the well-deformed
mid-shell nuclei in the \mbox{$Z \approx 100$}, \mbox{$N \approx 150$} region 
that are the object of detailed spectroscopic studies. By contrast, most 
of the heaviest recently synthesised nuclides and their daughter nuclei with 
\mbox{$Z \gtrsim 110$}, \mbox{$N \gtrsim 164$} are in a region of transitional 
nuclei~\cite{Cwi05a}, which have soft deformation energy surfaces and in 
some cases also exhibit shape coexistence. For those, fluctuations in shape
degrees of freedom or the mixing of coexisting shapes might have 
a significant impact on the ground state properties
\cite{Fle04a,Ben06a,Ben08a,Del10a,Pra12a}. Both can be incorporated
into the modelling in the framework of the generator coordinate method 
(GCM)~\cite{Egi04a,Ben08LH}.

Figures~\ref{fig:corr} and~\ref{fig:qa} present results that consider 
pairing and quadrupole correlations beyond the mean field as described 
in Ref.~\cite{Ben06a}. The former are treated by projecting the
 mean-field state on good particle number. 
The nuclear shapes are restricted to axially symmetric and 
reflection-symmetric configurations. For these, the description of 
quadrupole correlations beyond the mean field simplifies from a 
five-dimensional calculation to the treatment of one Euler angle and 
the intrinsic deformation $\beta_2$. 
The same Skyrme interaction SLy4 is used together with a density-dependent 
pairing interaction to generate the mean-field states and 
for the configuration mixing. 
A detailed analysis of the ground-state correlations obtained in such 
calculations for nuclei up to \mbox{$Z \approx 100$} has been
presented in Ref.~\cite{Ben06a}. These calculations were recently
extended to superheavy nuclei~\cite{Ben12SHE}. The resulting 
beyond-mean-field ground-state quadrupole correlation energies are 
displayed in Fig.~\ref{fig:corr}. Their behaviour is correlated to the 
ground-state deformation, Fig.~\ref{fig:beta2}, similar to what is found 
for the transition from the rare-earth to the lead region discussed in 
Ref.~\cite{Ben06a}. For the well-deformed nuclei below \mbox{$Z \lesssim 112$}
and \mbox{$N \lesssim 166$} the correlation energy is dominated by 
rotational energy from angular-momentum projection. Its value is 
on the order of 3~MeV and almost constant, and much smaller than the
static mean-field deformation energy that is on the order of about 
20~MeV. The correlation energy takes larger values up to about 4~MeV 
for the transitional systems with \mbox{$116 \lesssim Z \lesssim 124$} 
and \mbox{$168 \lesssim N \lesssim 184$}, and then drops rapidly to 
small values of about 1.2~MeV for the rigid spherical nuclei close to 
the double shell closure \mbox{$Z = 126$}, \mbox{$N = 184$}. For nuclei 
in the vicinity of spherical shell closures, the quadrupole correlation 
energy is on the same order or even larger than the static deformation 
energy. The variation of the correlation energy offers an explanation for 
the characteristic pattern of the measured two-nucleon separation energies 
across shell closures~\cite{Ben08a} in nuclei up to the Pb region, which 
is not described by pure mean-field calculations. Figure~\ref{fig:qa} 
illustrates its impact on $Q_{\alpha}$ values of very heavy nuclei. 
The difference between the mean-field and beyond-mean-field results is
evidently largest where the correlation energy displayed in 
Fig.~\ref{fig:corr} varies quickly. This happens around the 
\mbox{$N=184$} shell closure, where its variation from one
nucleus to its daughter in an $\alpha$-decay chain might be as
large as 3~MeV, and to a lesser extend for nuclei just below
\mbox{$Z=126$} and throughout the transitional region down to 
$Z$ values around 110. By contrast, for the well-deformed nuclei 
with \mbox{$Z \approx 100$}, the $Q_\alpha$ values are not 
significantly affected as the correlation energy is almost constant.
When drawing the $Q_\alpha$ values for isotopic chains as a function
of neutron number as in Fig.~\ref{fig:qa}, large gaps between
the curves correspond to large deformed or spherical proton shell 
effects, whereas large downward jumps of the curves correspond to 
neutron shell effects. Both can be easily associated with the 
appearance of large gaps in the single-particle spectra displayed
in Fig.~\ref{fig:spe:def:min}. Again, the empirical
shell effect at \mbox{$N=152$} is not correctly described. This 
is a common deficiency of virtually all self-consistent 
mean-field models. On the mean-field level, very similar results 
are obtained also with other Skyrme interactions \cite{Bur98a},
the D1s parametrisation of the Gogny force \cite{Del06a,War12a} and 
with relativistic mean-fields \cite{Bur98a,Ben00a}. It is noteworthy that 
the $Q_\alpha$ values from macroscopic-microscopic methods also 
underestimate the discontinuity at \mbox{$N=152$}, although they 
predict a sizable \mbox{$N=152$} gap in the single-particle
spectrum, cf.\ Ref.~\cite{Bar05a,Smo97a}. Curiously, the 
beyond-mean-field correlation energy has a visible impact on 
energy differences only around spherical shell closures, but 
not deformed ones. 

A similar method is the construction of a microscopic Bohr Hamiltonian 
through a series of approximations to projected GCM as described
in Refs.~\cite{Fle04a,Del10a}. The study of selected $\alpha$-decay chains
using a five-dimensional collective Bohr Hamiltonian in the full 
$\beta_2$-$\gamma$ plane derived from a relativistic mean-field 
has been recently presented in Ref.~\cite{Pra12a}.

For some of the nuclei discussed here, considering only axial and 
reflection-symmetric shapes might not be sufficient. For some 
of the transitional nuclei in the \mbox{$Z \approx 116$}, 
\mbox{$N \approx 170$} region, shallow triaxial minima have 
been predicted~\cite{Cwi05a}. Nuclei with neutron number 
slightly larger than $184$ might be reflection asymmetric 
in their mean-field ground states~\cite{War12a}, which also 
hints at a possible octupole softness of adjacent nuclei.
Likewise, based on the observation of low-lying $2^-$ levels at energies 
below two times the pairing gap it has been argued that 
nuclei around $N=150$ might be soft in non-axial octupole degrees 
of freedom \cite{Che08a}. 

An alternative for the description of correlations beyond the self-consistent 
mean field is the framework of particle-vibration coupling 
(PVC). There, correlations are added as a diagrammatic expansion 
in terms of single-particle degrees of freedom and RPA phonons. 
This methodology is currently restricted to spherical nuclei. A study 
addressing singly- and doubly-magic $Z=120$ isotopes in the relativistic 
mean-field framework \cite{Lit11a,Lit12a} finds that the single-particle
strength becomes highly fragmented, consequence of the small
gap in the unperturbed single-particle spectrum. The single-particle
levels are shifted toward the Fermi energy, such that the gap in 
the single-particle spectrum becomes smaller.
With 4-5~MeV, the PVC correlation energies of even-even \mbox{$Z=120$} 
nuclei reported in Ref.~\cite{Lit12a} are larger than the projected GCM 
values displayed in Fig.~\ref{fig:corr}. However, these values should 
not be directly compared, as the former include also correlation energy 
from pairing which is not contained in the latter.

%
%
\section{Conclusions}
\label{sect:conclusions}

To summarise the main points of our discussion
\begin{itemize}
\item
Because of the large density of single-particle levels, already small 
shifts of single-particle energies can qualitatively change the 
appearance of the single-particle spectra of SHE, which is not the 
case for lighter doubly-magic nuclei up to \nuc{208}{Pb}.
\item
For SHE, the importance of the spherical shell closures for magicity
is different than in lighter nuclei. Because of the 
presence of highly-degenerate high-$j$ levels and the smallness of 
the gaps in the single-particle spectrum, additional binding from 
quantal shell effects originates from the bunching of low-$j$ 
orbits near the Fermi energy, not from the gaps.
\item
It is not the magnitude of the shell effects that counts for the
stabilisation of SHE, but how quickly it varies with deformation.
One, sometimes even two, spherical shell closures do not
automatically prevent superheavy nuclei from having a deformed
mean-field ground state.
\item
Self-consistency induces many local structural changes in SHE.
\item
Their extrapolation to SHE amplifies differences between the models 
and parametrisations. 
\item
Almost all models agree that neutron shell closures enforce 
sphericity much more than proton shell closures do. This finding, 
however, has to be taken with a grain of salt. As pointed out in 
Refs.~\cite{Ben06a,Ben08a}, mean-field models tend to overestimate
the size of neutron shell gaps for shell closures throughout the
chart of nuclei up to $N=126$, whereas the proton shell closures
are usually more satisfactorily described.
\item
Quadrupole correlations beyond the mean field affect 
mass differences such as $Q_\alpha$ values in the transitional 
region between the well-deformed actinides and the spherical 
shell closures.
\end{itemize}

%
%

\ack

This research was supported in parts by 
the French \emph{Agence Nationale de la Recherche} under Grant 
No.~ANR 2010 BLANC 0407 "NESQ", 
the CNRS/IN2P3 through the PICS No.~5994,  
the PAI-P6-23 of the Belgian Office for Scientific Policy and by
by the European Union's Seventh Framework Programme ENSAR under 
grant agreement n262010.

%
%

\end{document}